\documentclass[11pt,aps,prd,nofootinbib,groupedaddress,preprintnumbers]{revtex4}
\usepackage{graphicx,epsfig}
\newcommand{\bea}{\begin{eqnarray}}
\newcommand{\beq}{\begin{equation}}
\newcommand{\eea}{\end{eqnarray}}
\newcommand{\eeq}{\end{equation}}

\newcommand{\nn}{\nonumber}
\newcommand{\Frac}[2]{\frac{\displaystyle{#1}}{\displaystyle{#2}}}
\newcommand{\lsim}{\raise0.3ex\hbox{$\;<$\kern-0.75em\raise-1.1ex\hbox{$\sim\;$}}}
\newcommand{\gsim}{\raise0.3ex\hbox{$\;>$\kern-0.75em\raise-1.1ex\hbox{$\sim\;$}}}
\newcommand{\eq}[1]{Eq.~(\ref{#1})}
\newcommand{\unity}{{\hbox{1\kern-.8mm l}}}

\newcommand{ \overdot}{{\raise .2 ex \hbox to 0pt {\hss\bf\smash{.}\hss}}}
\newcommand{\bpm}{\begin{pmatrix}}
\newcommand{\epm}{\end{pmatrix}}

\begin{document}

\preprint{FTUV-08-0626}
\preprint{IFIC-08-22}

\title{About a  (standard model) universe dominated by the right matter}
\author{G.~Barenboim}
\affiliation{Departament de F\'{\i}sica Te\`orica and IFIC, Universitat de 
Val\`encia-CSIC, E-46100, Burjassot, Spain.}
\author{O. Vives}
\affiliation{Departament de F\'{\i}sica Te\`orica and IFIC, Universitat de 
Val\`encia-CSIC, E-46100, Burjassot, Spain.}

\begin{abstract}
We analyze the phenomenology of a prolonged  early epoch of matter 
domination by an unstable but very long-lived massive particle.
This new matter domination era can help to relax some of the requirements
on the primordial inflation. Its main effect is the huge entropy production
produced by the decays of such particle that can dilute any possible unwanted 
relic, as the gravitino in supersymmetric models, and thus relax the 
constraints on the inflationary reheating temperature. A natural candidate 
for such 
heavy, long-lived particle already present in the Standard Model of the 
electroweak interactions would be a heavy right-handed neutrino.  
In this case, we show that its decays can also generate the observed baryon 
asymmetry with right-handed neutrino masses well above the bound from
gravitino overproduction.
\end{abstract}

\maketitle

\section{Introduction}
Inflation was introduced in the 80s \cite{inflation} as a solution to several
problems of the big bang cosmology. Perhaps the main problem was the large-scale
smoothness problem. Why different patches of the Universe,
that were not in causal contact in the radiation last scattering era, have
approximately the same  temperature today \cite{Hinshaw:2008kr}?. In inflationary models, an epoch of
exponential expansion inflates a small patch of the Universe in causal
contact to contain all
the observable Universe today. Simultaneously if the temperature after inflation is low enough, 
inflation helps also to 
dilute unwanted relics from higher scales and reduces the flatness problem.

It is usually assumed that some kind
of inflation starts already at the Planck scale to avoid the Universe collapse
in a few Planck times if $\Omega>1$ or (for any $\Omega$) to prevent the 
invasion of the surrounding 
inhomogeneity to our homogeneous patch before inflation.
On the other hand, the scales observable today   in the cosmic microwave background 
left the horizon at an energy $V^{1/4} \lsim 6 \times 10^{16} $ GeV \cite{Alabidi:2005qi}, or 60 
$e$-foldings 
before the end of inflation. So that, inflation must end below this scale.
 After the end of inflation
comes an era of reheating when the inflaton field oscillates around its minimum
and decays to ordinary particles. The final reheating temperature where we
recover ordinary big bang cosmology can take any value from $V^{1/4}$ above 
to scales as low as 1 MeV. However, the required dilution of unwanted relics,
as GUT monopoles or gravitinos in supersymmetric models, forces the
reheating temperature to be well below the GUT scale or even below $T_{RH}
\leq 10^{8}$ GeV in SUSY models.   

In this letter, we propose a simple and economic mechanism that helps 
solving some of these problems and reduces the requirements on the primordial
inflationary mechanism without further additions to the particle spectrum of
the Standard Model {\bf with right-handed neutrinos} \footnote{From now on, 
we call the Standard Model with right-handed neutrinos simply 
``Standard Model''}. After an initial inflationary epoch (still necessary to
reproduce the observed correlation on temperature fluctuations at large
scales) we assume our Universe is radiation dominated for a short period and
then enters a matter domination era due to the existence of a heavy long-lived
unstable particle that decays to radiation well-before nucleosynthesis, when
we  connect with usual cosmology. In the Standard Model, as we will show, this
role could be played by a heavy right-handed neutrino. In the literature, it
is well-known that late time entropy release can help to ameliorate some of
the problems of standard cosmology. However, so far most of these works have
only considered moduli fields in supersymmetric theories (see for instance 
\cite{Nagano:1998aa,Kawasaki:1999na,Kohri:2004qu,Pradler:2006hh,Kitano:2008tk}) and their real presence in nature could be considered more 
speculative than the existence of right-handed neutrinos. 

As we show below, this matter domination
mechanism, naturally embedded in the SM, is able to help primordial inflation
in several aspects. A long period of matter domination can reduce mildly the 
number of
e-folds before the end of inflation at which observable perturbations were
generated, relaxing this way flatness conditions on the inflationary potential.
Moreover, the large entropy production in the decay of
this particle  completely  dilutes any unwanted relics, eliminating the
constraint on the inflation reheating temperature. 
In this sense our matter domination epoch has the same advantages as thermal 
inflation \cite{Lyth:1995hj},
without resorting to yet another scalar field and /or scalar potential.

\section{Rewriting the history of the Universe}

Let's assume for a moment that at an early time a massive particle dominated 
the energy balance of the Universe by many orders of magnitude. How does the observed Universe feel
this new epoch?. Are there any observable consequences of this?

As it is well-known \cite{Kolb:1990vq,Dodelson:2003ft}, 
a massive particle $X$ becomes non-relativistic when the
temperature of the thermal bath falls bellow its mass, $M_X$, and its energy
density freezes out when it drops out of equilibrium. Then, $X$-relic
abundance relative to photons (radiation) becomes constant and therefore, if
$X$ is completely stable, the energy density of the $X$ particles will
eventually become larger than the radiation one, dominating the energy balance
of the Universe. If, $X$ is not completely stable but rather sufficiently
long-lived to dominate the energy density, later on $X$ will decay into
relativistic particles that thermalize. The radiation content of the
Universe will be increased and it will enter again in a radiation dominated
era. This will be the basic evolution of the Universe in our model.

We start from a radiation dominated Universe at a scale larger than the
mass of our $X$-particle, $M_X$, where this particle is in thermal
equilibrium. We assume that this particle decouples from the plasma at a
temperature of the order of its mass. This particle is unstable,
although very long-lived, i.e. it has very weak interactions with radiation
degrees of freedom. Then, its energy density, $ \rho_X$,  starts diluting as
matter, much 
slower than radiation. If its lifetime, $\Gamma_X$, is long enough, it will
necessarily dominate the energy density of the Universe. 
Although our $X$ particle decays all the time through   
an exponential law \cite{Turner:1984ff}, it is only when the age of the Universe is 
of the order of $1/\Gamma_X$ that the decay will sizeably reduce its
abundance. Once it reaches this point,  it will fastly decay into radiation
and our Universe  will go back to a radiation dominated  epoch  where it will
connect with the usual cosmology. This ``matching'' with the standard scenario
must happen well before nucleosynthesis.

The evolution equations for the matter and radiation energy-densities are 
well-known:
\bea
\label{eq:fulleqs}
\dot{\rho}_X &=& - 3 H (1+w_X) \rho_X - \Gamma_X \rho_X \\[.1cm]
\dot{\rho}_r^{\mbox{\small{old}}} & = & - 4 H \rho_r^{\mbox{\small{old}}} \\[.1cm]
\dot{\rho}_r^{\mbox{\small{new}}} & = & - 4 H \rho_r^{\mbox{\small{new}}} + \Gamma_X \rho_X  \\[.1cm]
H^2  & = & {\dot a \over a}~=~\frac{8 \pi}{3 M_{Pl}^2} \left( \rho_r^{\mbox{\small{old}}} +
\rho_r^{\mbox{\small{new}}} + \rho_X \right)
\label{eq:fulleqsb}
\eea
where $w_X$ is the equation of state parameter of particle $X$,
which drops from 1/3 to zero as $X$ becomes nonrelativistic, 
$\rho_r^{\mbox{\small{old}}}$ is the energy density in radiation not related
to $X$ decays, $\rho_r^{\mbox{\small{new}}}$ is the energy density in radiation
produced by $X$ decays and here we assume a flat Universe after inflation
consistent with observations \cite{Hinshaw:2008kr}.
 
From these equations we can see that, when $X$ is nonrelativistic, 
the number of $X$'s per comoving volume
($N_X = R^3 \rho_X/M_X$) follows a simple exponential decay law and the 
(formal) solution to these equations is given by:
\bea
\label{eq:fulleqs2}
\rho_X &=& \rho_X^0 \left[{ a \over a^0 }\right]^{-3} ~ e^{-\Gamma_X \left(t-t_0\right)}\\[.1cm]
{\rho}_r^{\mbox{\small{old}}} & = & {\rho_r^{\mbox{\small{old}}}}^0 \left[{ a \over a^0 }\right]^{-4}\\[.1cm]
\rho_r^{\mbox{\small{new}}} & = & \rho_X^0 \left[{ a \over a^0
  }\right]^{-4} \int^t_{t_0}  d t^\prime~\left[{ a(t^\prime) \over a^0}\right]
e^{-\Gamma_X t^\prime} \Gamma_X \\[.1cm]
H^2  & = & {\dot a \over a}~=~\frac{8 \pi}{3 M_{Pl}^2} \left( \rho_r^{\mbox{\small{old}}} +
\rho_r^{\mbox{\small{new}}} + \rho_X \right)
\eea
where the superscript zero denotes the value of that quantity at the initial
epoch.

\begin{figure}

\centerline{\epsfxsize 3.75 truein \epsfbox {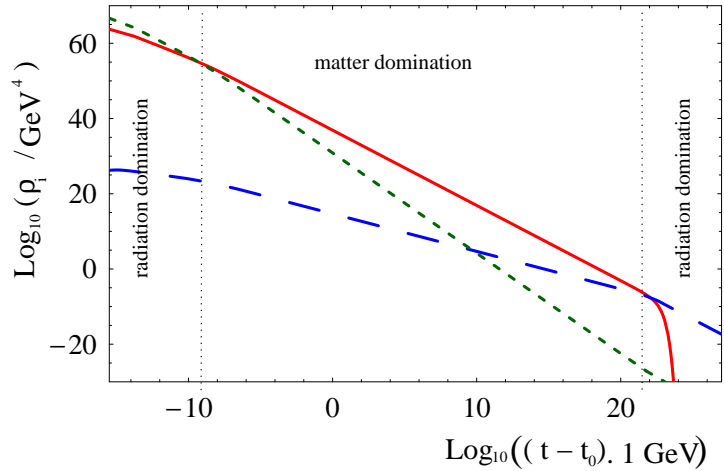} }
 
\caption{ Evolution of the different components of the energy density of the
  Universe from $T \sim 10^{14}$ GeV  to $T \sim 1$ MeV. Short-dashed (green) line corresponds to
  ${\rho_r^{\mbox{\small{old}}}}$ versus time, long-dashed (blue) line to 
${\rho_r^{\mbox{\small{new}}}}$ and the solid (red) line to $\rho_X$. We start
from $\rho_X= {\rho_r^{\mbox{\small{old}}}}/200$ at $T \sim 10^{15}$ GeV, with
a $\Gamma_X = 10^{-20}$ GeV.\label{fig:densities}
}
\end{figure}

\begin{figure}

\centerline{\epsfxsize 3.75 truein \epsfbox {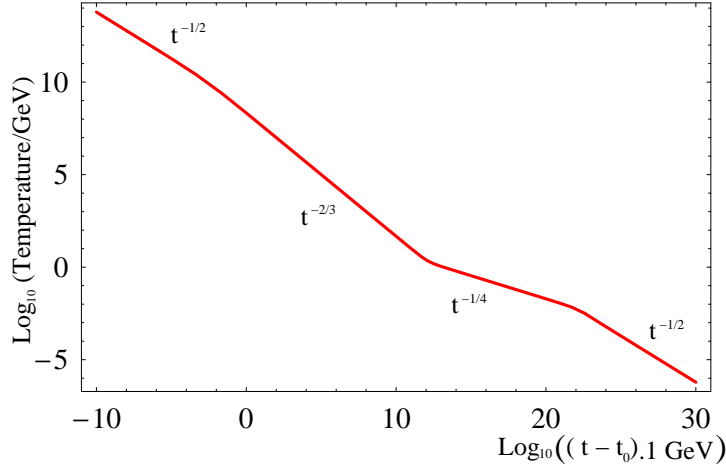} }
 
\caption
{Temperature of the Universe (in GeVs) as a function of time. The four
  different epochs in the evolution  of the Universe, radiation domination, matter domination,
  decay and radiation domination again can be seen in the different slopes of
  the curve.  The time dependence of the temperature is explicitly indicated for each epoch.
\label{fig:temperature}}
\end{figure}

In general, it is not possible to integrate analytically these equations,
although useful approximations exist \cite{Scherrer:1984fd,Turner:1984ff}.
However it is always possible to solve them numerically as we have done to 
generate Figures \ref{fig:densities} and \ref{fig:temperature}.
These evolution equations,  (\ref{eq:fulleqs}) to (\ref{eq:fulleqsb}),   
were thoroughly analyzed by M.~Turner and collaborators in
Refs.~\cite{Scherrer:1984fd,Turner:1984ff} (for a more recent work
see for example  \cite{Boubekeur}) and we agree completely with their
analysis of the matter and radiation densities, entropy and
temperature. However, we are specially interested in the particular limit of
small $\Gamma_X$ and large $M_X$. In fact, as we will see below, we will 
focus on the limit of small $\Gamma_X$ keeping 
$ \Gamma_X \gg  1/t_{BBN}$ so that no trace of $X$ is present at
nucleosynthesis time, in agreement with observations.

In this limit,
our massive particle $X$ dominates the energy density of the Universe by many
orders of magnitude during a sizeable fraction of the thermal history of the Universe 
(see Figure~\ref{fig:densities}). 
As shown in Refs.~\cite{Scherrer:1984fd,Turner:1984ff}, during
the $X$ decays the temperature of the Universe does not fall as $t^{-1/2}$ $(a^{-1})$, but
rather as  $t^{-1/4}$ $(a^{-3/8})$  due to the entropy release of the decays and
the temperature reaches an almost flat plateau from the point where the 
energy density in new radiation born through $X$ decays and the energy density 
in old radiation become comparable
up to $t\simeq \Gamma_X^{-1}$. After this time, $X$ rapidly decays and the 
temperature falls again as  $t^{-1/2}$ $(a^{-1})$ (see 
Figure~\ref{fig:temperature}). 

Once it has completely decayed away, our Universe is left with a temperature
\bea
\label{eq:temp}
T_{\mbox{\small{post-decay}}} \;\simeq  \;1.0 \cdot 10^9 \; \left( \frac{g_*}{200}\right)^{-1/4} 
\left( \frac{\Gamma}{1 \; \mbox{\small{GeV} }}\right)^{1/2}   \mbox{\small{GeV} }
\eea
where $g_*$ counts the effective number of relativistic degrees of
freedom. Notice that the temperature after  $X$-decay depends only on the
decay width $\Gamma$. The ratio of entropy per comoving volume before and 
after $X$ decay is given by
\bea
\label{eq:entropy}
\frac{S_{\mbox{\small{post-decay}}}}{ S_{\mbox{\small{pre-decay}}}}
 \;\simeq  \; 0.14 \; r \; \left( \frac{g_*}{200}\right)^{-3/4}
\left( {\frac{1 \;  \mbox{\small{GeV}}} {\Gamma}} \right)^{1/2} 
\left( \frac{M_X}{10^{10} \; \mbox{\small{GeV} }}\right) 
\eea
where $r= g_X/2$ if $X$ is a boson and  $r= 3 \, g_X/8$
if it is a fermion, with $g_X$ the total number of spin degrees of freedom.
\begin{figure}
\centerline{\epsfxsize 4.75 truein \epsfbox {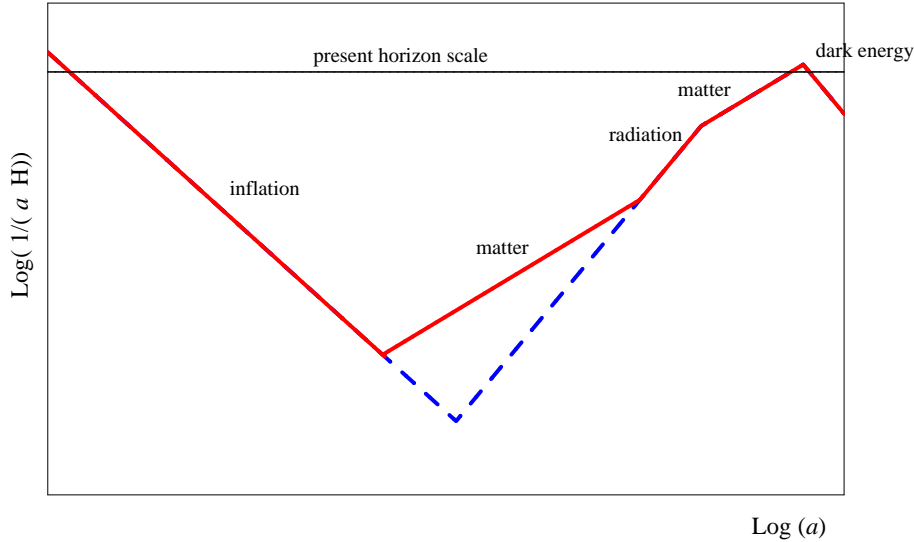} }
 
\caption
{Comoving Hubble radius, $\log \left(1/a H \right)$,
 versus $\log a$. This plot shows the different eras entering the
$e$-foldings calculation. Inflation is an epoch where $\log \left(1/a H \right)$ is 
decreasing. Exponential inflation gives a line with a slope of -1. In all other cases
the inflation line is shallower. During matter domination $\left(1/a H \right) \propto a^{1/2}$,
while during radiation domination  $\left(1/a H \right) \propto a$. The current dark energy 
domination signals a new inflationary epoch. The horizontal (black) solid line indicates
the present horizon scale. The number of $e$-foldings before the end of inflation at which
observable perturbations were born is the horizontal distance between the time when
$\left(1/a H \right)$ first crosses that value and the end of inflation.
The solid (red) line represents a Universe with a period of matter domination  before BBN. The
dashed (blue) line represents the standard cosmological history of the Universe with only one
(recent) epoch of matter domination. 
\label{fig:deltan}}
\end{figure}

As mentioned before, an early period of matter domination, triggered by a long lived massive particle
that goes out of equilibrium and comes to dominate  the energy density of the Universe before 
decaying, reduces the number of $e$-foldings before the end of inflation at which our
present Hubble scale equaled the Hubble scale during inflation, i.e. the time of horizon crossing.
The reduction is given by
\bea
\Delta N  = \frac{1}{12} \ln 
\left( \frac{\rho_{\mbox{\small{post-matter-d}}} }{\rho_ {\mbox{\small{pre-matter-d}}}}\right) 
\eea
where $\rho_{\mbox{\small{pre-matter-d}}}$ and  $\rho_{\mbox{\small{post-matter-d}}}$ 
are the energy densities  at the beginning and end of the matter dominated era, respectively.
The reduction can also be expressed in terms of the $X$ parameters as
\bea
\Delta N  \;\simeq \;-\frac{1}{6} \ln 
\left( 90 \;  g_*^{-3/2} \;  r^2  \;  \frac{M_X^2 }{\Gamma M_{Pl}}\right) 
\eea
This reduction is illustrated in Figure~\ref{fig:deltan}, where it can be
clearly seen that to determine the number of $e$-foldings after horizon
crossing of a given cosmological scale, as the present Hubble scale, 
the complete thermal history of the Universe must be used.  
From nucleosynthesis onwards this
history is well in place. However earlier epochs are still very uncertain. The
standard cosmological model assumes that inflation gives way to a long period
of radiation domination (we neglect here the period of reheating that
immediately follows inflation and assume sudden transitions between the
different regimes).  The radiation dominated epoch lasts until a redshift of a
couple of thousands before entering an era of matter domination, which at
redshift below one gives way to the current acceleration.

Changing the sequence of events after inflation can therefore have a strong
impact on the the number of $e$-foldings calculation. If our Universe goes
through a long period of a regime where $\log \left(1/a H \right)$ scales as
$a^{n}$, i.e. $H \propto a^{-(n+1)}$, it is straightforward to see that with
$n > 1$ the total number of $e$-foldings will be increased while for $n < 1$ 
this number will be reduced.  A period of matter domination belongs to the 
latter class, as in a matter dominated epoch $\left(1/a H \right) \propto
a^{1/2}$ opening the door to a significant reduction on the number of
$e$-foldings. 

If we put all this information together we can see what are the required 
features of our particle $X$, its mass $M_X$ and its lifetime $\Gamma $,
if it is to dominate over the energy density of the Universe for a long period.

Although we do want a prolonged period of matter domination, we want it to come
to an end  {\bf at the latest} shortly before nucleosynthesis, as the Universe
must have attained thermalized radiation domination by that time. Using 
Eq.~(\ref{eq:temp}), this condition sets a lower bound on $\Gamma $,
\bea
\Gamma \geq 2.0 \cdot 10^{-24}  \; \left( \frac{g_*}{200}\right)^{1/2} 
\;\mbox{\small{GeV} }.
\eea
We can also get un upper bound on  $\Gamma $, by requesting
the reheating temperature $ T_{\mbox{\small{post-decay}}}$ to be at most
$10^8$ GeV, so that no unwanted relics will be produced after $X$ decay in
supersymmetric models. Such a condition reads
\bea
\Gamma \leq 0.7 \; \left( \frac{g_*}{200}\right)^{1/2} 
\;\mbox{\small{GeV} }.
\eea
However, for lifetimes this long,  we can see from  \eq{eq:entropy} 
that only large X-masses are capable of effectively diluting the
unwanted relics. In the case of the gravitino this even more difficult as the
gravitino abundance is also proportional to the temperature, $T \geq M_X$. 

Of course detailed bounds can be set only after specifying the basic physics 
behind $X$, its production mechanism, its decays, or in short, its
interactions.  Nevertheless, we can say that requiring at 
least five orders of magnitude of dilution by entropy production for 
$M_X \le 10^{10}$ GeV would need  
\bea
2.0 \cdot 10^{-6 } \; \left( \frac{g_*}{200}\right)^{1/2} \;\mbox{\small{GeV} 
} \geq \Gamma \geq 2.0 \cdot 10^{-24} \; \left( \frac{g_*}{200}\right)^{1/2}
\;\mbox{\small{GeV} }.  
\eea 
In this case the five orders  of magnitude of increase in the entropy 
should be enough to get rid of unwanted relic which could have been produced 
at earlier times for $ T_{\mbox{\small{post-decay}}} \lsim 10^8$ GeVs. Larger 
lifetimes are also possible if we do not need such a large entropy
production.  

With regards  to the reduction in the number of $e$-foldings, only a very
prolonged period of matter domination, i.e. large $M_X$ and $\Gamma$ in the
lower part of  the allowed range, is required to give a significant reduction.
In most cases, however, this number is expected to be below 10.

\section{Matter domination in the Standard Model}
 
Now, we must check whether this early matter domination epoch could exist in
the context of the Standard Model of the strong and electroweak interactions or
any of its extensions. Clearly to obtain such an early matter domination era
we  need a massive particle, $X$, in thermal equilibrium at temperatures above
its mass with a very long lifetime.
The  simplest candidate for our $X$-particle would be a right-handed neutrino.
Right-handed neutrinos are one of the minimal additions to the Standard Model
to  reproduce the observed neutrino masses through the seesaw mechanism \cite{seesawrev}. 
These
right-handed  neutrinos, $R_i$, have super-heavy masses, that can be as high 
as the Grand Unification scale, and are singlets under the SM gauge group. 
The only  renormalizable couplings of $R_i$ with the SM particles are, 
possibly small, Yukawa couplings. Therefore, if these couplings are 
sufficiently  small, it seems possible that the right-handed neutrinos play the
role of $X$-particle with a long lifetime.

More precisely,  the right-handed neutrino masses and Yukawa couplings have to
reproduce the measured neutrino masses and mixings though the seesaw
mechanism,  $m_{\nu_L} = v_2^2~Y_{\nu}\cdot\left(M_R\right)^{-1}\cdot
Y_{\nu}^T$, with $v_2$ the vacuum expectation value of the up-type Higgs.  
From here it  is straightforward to obtain the required
right-handed Majorana matrix from the seesaw formula itself: 
\bea 
\label{invseesaw}
M_R =  v_2^2~Y_{\nu}^T\cdot\left(m_{\nu_L}\right)^{-1}\cdot Y_{\nu}
\eea
From the  light neutrino mass matrix, $m_{\nu_L}$, we know the mixings and the
two mass differences. The mixing matrix $U$ is close to the so-called
tribimaximal  mixing, 
\bea
\label{Umix} 
U=\begin{pmatrix}{ \frac{2}{\sqrt{6}} & \frac{1}{\sqrt{3}} & 0 \cr-\frac{1}{\sqrt{6}} &\frac{1}{\sqrt{3}}
&-\frac{1}{\sqrt{2}} 
\cr-\frac{1}{\sqrt{6}} &\frac{1}{\sqrt{3}}& \frac{1}{\sqrt{2}}} \end{pmatrix}.
\eea 
Then, $m_{\nu_L} = U^*\cdot \mbox{Diag}\left(m_1,m_2,m_3\right) \cdot U^\dagger$,
and the  inverse of this matrix is $m_{\nu_L}^{-1} = U\cdot 
\mbox{Diag}\left(\frac{1}{m_1},\frac{1}{m_2},\frac{1}{m_3}\right)
\cdot U^T$.  Therefore,
\bea
\label{connect1} 
m_{\nu_L}^{-1} =& \Frac{1}{m_3} \bpm {0&0&0\cr 0&\frac{1}{2}&-\frac{1}{2}\cr
  0&-\frac{1}{2}&\frac{1}{2} }\epm + \Frac{1}{m_2} \bpm 
{\frac{1}{3}&\frac{1}{3}&\frac{1}{3}\cr \frac{1}{3}&\frac{1}{3}&\frac{1}{3}\cr
  \frac{1}{3}&\frac{1}{3}&\frac{1}{3}}\epm + \Frac{1}{m_1}  \bpm
{\frac{2}{3}&-\frac{1}{3}&-\frac{1}{3}\cr
  -\frac{1}{3}&\frac{1}{6}&\frac{1}{6}\cr 
  -\frac{1}{3}&\frac{1}{6}&\frac{1}{6}} \epm.& 
\eea  
Experimentally we have
that  $m_3 = m_{\rm atm} \simeq 0.05$ eV, $m_2= m_{\rm sol} \simeq 0.008$ eV
and $m_1 \ll m_2$ in the normal hierarchy situation and $m_2 = m_{\rm atm}
\simeq 0.05$ eV,  $m_1= m_{\rm atm} - m_{\rm sol}/2 \simeq 0.046$ eV and $m_3
\ll m_2$ in the inverse hierarchy case \cite{neutrinorev}.
 
As seen in  \eq{invseesaw} the masses of the right-handed neutrinos reproducing
the observed light-neutrino masses and mixings are determined by the Yukawa
matrix,  $Y_{\nu}$. Choosing the basis of diagonal $Y_{\nu}^\dagger Y_{\nu}$
and diagonal charged lepton Yukawa matrix, we have $Y_{\nu} = V_L\cdot
\mbox{Diag}\left(y_1,y_2,y_3\right)$.  Obviously the physics depends strongly
on the form of $Y_{\nu}$, both on the eigenvalues, $y_i$, and the $V_L$
matrix. Let us 
first analyze the role of $V_L$. We have two limiting situations: a) $V_L$ has
large mixings and is the source of the observed PMNS matrix in neutrino
mixings, $V_L\simeq U^*$ and b) the mixings in $V_L$ are small, similarly to 
the situation observed in the CKM mixing matrix, $V_L\simeq \unity$.

Case a) is very simple, the seesaw mechanism plays no role in the generation
of the neutrino mixings. The observed large neutrino mixings are already
present in the Yukawa couplings before the seesaw mechanism. This corresponds 
to the situation where, the light neutrino Majorana mass matrix and the
neutrino Yukawa couplings, or equivalently, 
the right-handed neutrino Majorana matrix and the 
Yukawa combination $Y^\dagger Y$, can be simultaneously diagonalized.  
So, we have $M_R = v_2^2~
\mbox{Diag}\left(y_1^2/m_1,y_2^2/m_2,y_3^2/m_3\right)$. The decay widths of
the right-handed neutrinos will be given by $\Gamma_i = \frac{1}{8 \pi} M_i 
\left(Y^\dagger Y\right)_{ii}$ and we must
compare it with the Hubble rate, $H (T=M_i)$ in order to know if/when our
massive neutrino will go out of equilibrium. Equivalently, we can compare the
effective mass $\tilde m_i = \left(Y^\dagger Y\right)_{ii} v^2 /M_i$
(i.e. $\Gamma = \frac{\tilde m_i}{8 \pi}~\frac{\tilde M_i^2}{v_2^2}$) with the
critical mass $m_* = 1 \times 10^{-3}$ eV
\cite{Fischler:1990gn,Buchmuller:1992qc}.  A right-handed neutrino would
dominate the energy density if $\tilde m_i < m_*/g_*$ where $g_*$ is the number
of radiation degrees of freedom at $T=M_i$.  
The presence of $g_*$ is due to the fact that in a time 
$H^{-1}$ we can see from \eq{eq:fulleqs2} that the ratio of matter and radiation
densities grows as $a$. But matter has to overcome $g_*$ radiation degrees of 
freedom and hence to dominate the energy density it needs a lifetime $g_*$ 
times longer\footnote{However, as we see below,
  this large effective mass, $\tilde m_i \simeq 10^{-5}$ eV, does not generate 
sufficient entropy.}. Altogether, in case a), it is clear 
$\tilde m_3 =m_3$, $\tilde m_2=m_2 $ and  $\tilde m_1 = m_1$. 

Therefore, in the normal hierarchy case, taking $m_1 \leq 10^{-10}$ eV, $R_1$ 
would dominate the energy density of the Universe with a mass of, 
\beq
\label{eq:MR1} 
M_{R_1} = \left( \Frac{y_1}{10^{-6}}\right)^2 \left( \Frac{1 \times 10^{-10}
    \mbox{eV}}{m_1}\right)  6 \times 10^{11}~ \mbox{GeV}.
\eeq  
This case is therefore a perfect example of how a right-handed neutrino can 
dominate the energy density of the Universe after inflation. In terms of
$\tilde m_i$ we can write \eq{eq:entropy} as 
\bea
\label{eq:entropym} 
\frac{S_{\mbox{\small{post-decay}}}}{ S_{\mbox{\small{pre-decay}}}} \;\simeq
\; 0.4 \; r \; \left( \frac{g_*}{200}\right)^{-3/4} \left( {\frac{1 \times 
      10^{-6} \; \mbox{\small{eV}}} {\tilde m_i}} \right)^{1/2} 
\eea 
Therefore, if we want two orders of magnitude of entropy production, we would
need $m_1 = 10^{-10}$ eV (corresponding to $\Gamma = 4.6 \times 10^{-2}$ GeV) 
and $T_{\mbox{\small{post-decay}}} \;\simeq 2 \times 10^8$ GeV. Here, as the
gravitino abundance is approximately linear with the reheating temperature
\cite{Kawasaki:2004qu}, two 
orders of magnitude of dilution by the entropy production would
correspondingly relax 
the bound on the reheating temperature by two orders of magnitude. Naturally,
this can be easily improved by choosing smaller $m_1$ and $y_1$ in 
Eqs.~(\ref{eq:MR1}) and (\ref{eq:entropym}).

However this 
simple model has several phenomenological problems. First, given that $\tilde
m_1\ll m_* $,  this right-handed neutrino would not be produced thermally
through its Yukawa interactions. In fact this will be a common problem of any
massive particle  dominating the energy density of the universe as necessarily
its decay/production rate will be much slower than the Hubble rate. Therefore
we will always need  another active interaction to produce our right-handed
neutrino in the thermal plasma after inflation. This role could be played, for
instance, by a gauged B--L interaction. Many Grand Unified models based
on $SO(10)$ or groups containing it have an intermediate scale of the order of
$10^{13}$ GeVs with a intermediate gauge group containing $U(1)_{B-L}$, as 
$SU(2)_L\times SU(2)_R\times SU(3)_c\times U(1)_{B-L}$ for example\cite{Sato:1995du,Aulakh:1999pz,Aulakh:2000sn}.
In these grand unified models the B--L coupling unifies with the other gauge
couplings at $M_{GUT}$ and therefore it is always strong enough to keep the 
right-handed neutrinos in thermal equilibrium  
in the unbroken phase. However, the B--L gauge interaction can never mediate 
the neutrino decay as it couples only diagonally in flavour. The neutrino
decay would require a Yukawa interaction to lighter states that, as shown
above, in this case is very small.
A second problem, more specific of this
 particular case is that the decay of this right-handed neutrino erases 
completely any previously existing baryon or lepton asymmetry and therefore, 
 we need some mechanism to generate the observed baryon asymmetry. The baryon
asymmetry generated by this completely out-of-equilibrium decay of the 
 right-handed neutrino is given  by \cite{baryo}
\bea  
\label{epsilon}
\eta_{\rm B} =   \Frac{8}{23}~ \varepsilon \simeq ~\Frac{1}{16 \pi} ~ 
\sum_{j\neq 1} \Frac{{\rm Im} \left[ \left(Y^\dagger Y\right)_{1 j}^2\right]}{\left(Y^\dagger Y\right)_{1 1}} ~ \Frac{~M_1}{~M_j},
\eea    
but, as $Y^\dagger Y$ is completely diagonal in the basis
of diagonal  right-handed neutrino masses, no new lepton asymmetry is
generated by $R_1$ decays. 

However, this 
situation is very unstable and a slight departure from the perfect alignment
of the Majorana and Yukawa matrices changes the situation. If we call $R$ the 
rotation diagonalizing the neutrino Yukawas in the basis of diagonal
left-handed neutrino Majorana masses, we have $Y_{\nu} = U^*\cdot R \cdot  
\mbox{Diag}\left(y_1,y_2,y_3\right)$ and, 
\bea 
\left(M_R\right)_{i j} = v_2^2~y_i y_j \left(\frac{1}{m_1}~ R_{1 i}R_{1j}~ +~
\frac{1}{m_2}~ R_{2 i}R_{2 j}~ +~\frac{1}{m_3}~ R_{3 i}R_{3 j}\right).
\eea 
If
the Yukawa couplings are sufficiently hierarchical, $y_3 \gg y_2\gg y_1$, the 
 heaviest eigenvalue will be given by  the
element  $\left(M_R\right)_{3 3}$, 
\bea 
\label{invseesawR} 
\left(M_R\right)_{3 3}  \simeq v_2^2~y_3^2 \left(\frac{1}{m_1}~
(\sin \theta_{13})^2 +~
  \frac{1}{m_2}~(\sin  \theta_{23})^2~ +~\frac{1}{m_3}~
  (\cos\theta_{13}\cos\theta_{23})^2 \right) \, ,
\eea    
where we used the standard PDG parametrization for the matrix $R$ \cite{PDG}. 
From here,  we can see that the contribution from $m_1<< m_2,m_3$, will dominate
$\left(M_R\right)_{3 3}$, and hence the heaviest right-handed neutrino
eigenvalue,   if $(\sin \theta_{13})^2 >
m_1/m_3$. For $m_1 = 1 \times 10^{-10}$ eV, a $\sin \theta_{13} > 4.4 \times
10^{-5}$ will be enough  and such a small departure from perfect alignment
will completely change the situation.
To understand this, 
it is enough to analyze a simpler situation with $\theta_{12}=\theta_{23}=0$
and $\theta_{13}\neq 0$.  Let us take $y_3 \simeq 1$, $y_1 \simeq 10^{-6}$ 
(similar to the up-quark hierarchy), $m_3 \simeq 0.05$ eV and 
$m_1 \simeq 10^{-10}$ eV.  In this case, the two right-handed neutrino
eigenvalues (the other one is unchanged) are given by,
\bea 
\label{Reigenv}
M_{R_3}   &\simeq& v_2^2 y_3^2 ~\left( \Frac{\cos^2 \theta_{13}}{m_3} + 
\Frac{\sin^2 \theta_{13}}{m_1} \right), \nn \\
M_{R_1}   &\simeq& v_2^2 y_1^2~ \Frac{1}{ m_1~ \cos^2 \theta_{13} + m_3~ \sin^2
    \theta_{13}}. \nn \\
\eea  
If $\sin \theta_{13} \gg m_1/m_3$ then both $M_3$ and $\tilde m_3$ are fixed 
by $m_1$,  while $M_1$ and $\tilde m_1$ are fixed by $m_3$. So, we have,
for $m_1$ in the interesting range:
\bea 
\label{MR3mix}
M_{R_3}& =& \left( \Frac{y_3}{1}\right)^2 \left( \Frac{1 \times 10^{-10}
    \mbox{eV}}{m_1}\right) \left( \Frac{\sin \theta_{13}}{0.005}\right)^2 1.5
\times 10^{19}~ \mbox{GeV},\nn \\
\tilde m_3 &=& m_1/\sin^2 \theta_{13}
\eea  
And this is the only neutrino that can dominate the energy
density of the Universe. Clearly, this situation is not interesting  
phenomenologically as it is not possible to produce it after inflation and
(probably) it does not produce a large amount of entropy.  
 
The most interesting situation corresponds to  $\sin \theta_{13} < m_1/m_3$. In
this  case  from \eq{Reigenv} we have $M_{R_3} \simeq v_2^2~ y_3^2/m_3 \left( 1 +
    \sin^2\theta_{13}~m_3/m_1\right)$ and $M_{R_1} \simeq v_2^2~ y_1^2/\left(m_1
   \left( 1 + \sin^2\theta_{13}~m_3/m_1\right)\right)$. Now, the rotation that
  diagonalizes the right-handed mass matrix is given by $\sin \phi \simeq \sin
 \theta_{13} ~ (y_1 m_3)/(y_3 m_1)$. Therefore in the basis of diagonal
  right-handed neutrino masses  we have,
 \bea
\label{YYrot} 
Y^\dagger Y =\begin{pmatrix}{y_1^2 \left( 1 + \sin^2 \theta_{13}
     \left(\Frac{m_3}{m_1}\right)^2\right) & 0 & - y_1 y_3 ~\Frac{m_3}{m_1}~
\sin \theta_{13} 
\cr 0   & y_2^2 & 0
\cr - y_1 y_3 ~\Frac{m_3}{m_1}~\sin \theta_{13} & 0 & y_3^2 }\end{pmatrix},
\eea   
and for $\sin \theta_{13} < m_1/m_3$ we have $\tilde m_1 \simeq m_1$ and 
$\tilde m_3 \simeq m_3$.  This means the lightest right-handed neutrino, with
a mass approximately given by \eq{eq:MR1}, can
still dominate the energy density.  
In such a situation, we can  see from \eq{epsilon}, that the generated baryon 
asymmetry is given by  
\bea 
\label{goodeps} 
 \eta_{\rm B} \simeq ~\Frac{1}{16 \pi} ~ \Frac{{\rm Im} \left[ \left(y_1 y_3
\Frac{m_3}{m_1}~\sin \theta_{13} \right)^2\right]}{\left(y_1^2\right)} ~
\Frac{y_1^2 ~m_3}{y_3^2~m_1}  \simeq \Frac{1}{16 \pi}~ y_1^2
~\sin^2 \theta_{13}~ \Frac{m_3^3}{m_1^3}.  
 \eea         
Taking $y_1 \simeq 10^{-7}$, $m_3 \simeq 0.05$ eV, $m_1 \simeq 10^{-12}$ eV
 and $\sin \theta_{13}= 0.1 \times m_1/m_3$, we would obtain $\eta_{\rm B}
\simeq 10^{-7} \sin \varphi$ with  $\varphi$ the CP violating phase of 
 $(Y^\dagger Y)_{13}^2$. Therefore, in this case, it would be possible to
generate the observed baryon asymmetry  and \underline{simultaneously} dilute
the relic density and in particular the gravitino density by three orders of
magnitude. This means that the bound on the inflationary reheating temperature 
would be relaxed by three orders of magnitude. Clearly, using smaller values
for $m_1$ and $y_1$ the situation can be improved arbitrarily.
  
The second example of Yukawa mixing matrix was case b) where $V_L \simeq
\unity$ so that  we can neglect this rotation on $m_{\nu_L}$ as a small 
rotation will not modify the contribution of the different neutrino
eigenvalues to the matrix elements. Then, we have:
\bea
\label{right1}
M_R =& 
v^2 ~\bpm {y_1^2\frac{m_2^{-1}+ 2 m_1^{-1}}{3} &y_1 y_2\frac{m_2^{-1}- m_1^{-1}}{3}&y_1 y_3\frac{m_2^{-1}- m_1^{-1}}{3}\cr
y_1 y_2\frac{m_2^{-1}- m_1^{-1}}{3}&y_2^2 \frac{m_1^{-1}+ 2
  m_2^{-1}+ 3 m_3^{-1} }{6}&y_2 y_3\frac{m_1^{-1}+ 2
  m_2^{-1}- 3 m_3^{-1}}{6}\cr
y_1 y_3\frac{m_2^{-1}- m_1^{-1}}{3}&y_2 y_3\frac{m_1^{-1}+ 2
  m_2^{-1}- 3 m_3^{-1}}{6}&y_3^2 \frac{m_1^{-1}+ 2
  m_2^{-1}+ 3 m_3^{-1}}{6} }\epm.&  
\eea
We must diagonalize this matrix to obtain the right-handed neutrino
eigenvalues  and the Yukawa matrix in the basis of diagonal $M_R$.
In analogy with the charged lepton and quark Yukawas we can expect $y_3 \gg
y_2 \gg y_1$. Then we obtain, in the normal hierarchy case, 
\bea
M_{R_3}\simeq v^2~ y_3^3 ~ \frac{m_1^{-1}}{6} \;\;\;\;, \;\;\;\;
M_{R_2}\simeq  v^2~ y_2^3~ 2 m_3^{-1}\;\; \;\; \mbox{and}\;\;\;\;  
M_{R_1}\simeq  v^2~ y_1^3~ 3 m_2^{-1}.
\eea
Then, we have, 
$\tilde m_3 \simeq 6~ m_1$, $\tilde m_2 \simeq m_3 $ and  $\tilde m_1 \simeq \frac{2}{3}~
m_2$. This means that once again the right-handed neutrino that could dominate
the energy density of the Universe is the heaviest one and its  mass
would be given by \eq{MR3mix} with  $\sin \theta_{13} \sim 1/\sqrt{6}$, i.e. a
mass close or even above the Planck scale, unless $y_3$ is much smaller than 1.

Thus we see that with hierarchical Yukawa eigenvalues, similar to the up-quark
eigenvalues, it is possible to have right-handed neutrino dominance of the
energy density with consistent phenomenology, although only in rather
fine-tuned situations where $V_L$ is very close to the PNMS mixing matrix but
not exactly equal. If we move to an extension of the minimal Standard Model
 with right-handed neutrinos (or Minimal Supersymmetric Standard Model) the 
same results can be obtained without such a tight fine-tuning.   

From \eq{invseesaw}, we
can see that besides the Yukawa mixing $V_L$ we can still use different 
Yukawa eigenvalues. In a Grand Unified Theory (GUT)
with an underlying Pati-Salam symmetry,
we would expect the neutrino Yukawa couplings to be related to the up quark
Yukawas, and in fact we would expect one of the neutrino eigenvalues of the
order of the top Yukawa coupling \cite{Masiero:2002jn}. However the two light Yukawa eigenvalues
are less restricted. The masses of the right-handed neutrinos will depend on
the Yukawa eigenvalues and we can make them as small as we wish. However, if
$y_3$ is large, we will normally be in the same situation as before and the 
only neutrino with a sufficiently large lifetime will still be $\nu_{R_3}$,
which will be far too heavy. An interesting limit is when one of the Yukawa
eigenvalues is exactly zero, $y_1=0$, and therefore, one of the light
left-handed neutrino masses is zero.  Again the simplest situation is when
the right-handed Majorana matrix and the Yukawas, $Y^\dagger_\nu Y_\nu$ are
simultaneously diagonalizable. In this case, it is clear that only two 
right-handed neutrinos will
play a role in the seesaw mechanism and the third one will be completely
decoupled from the seesaw. In fact this third neutrino does not couple to the
doublets through Higgs Yukawa couplings. Given that right-handed neutrinos are
singlets under the SM group, apparently these neutrinos do not decay at all.
However, if we have a GUT symmetry, as for instance $SO(10)$ or a group
containing $SU(2)_R$, at a high scale, the right-handed neutrino will decay
with a lifetime,
\bea
\Gamma_{\nu_R} \simeq \alpha_{GUT}^2 \Frac{M_{R_i}^5}{M_{GUT}^4} \simeq
1\times 10^{-18} \left(25~\alpha_{GUT}\right)^2 \left(\frac{M_{R_i}}{10^{10}~
{\rm GeV}}\right)^5 
\left(\frac{2 \times ~10^{16}~{\rm GeV}}{M_{GUT}}\right)^4 {\rm~ GeV}.
\eea     
So, indeed we can see that this right-handed neutrino can dominate the energy
density of the Universe if it is produced through another interaction, as a
gauged B--L. In this case the production of the baryon asymmetry is not
possible through gauge interactions. However, it is possible that this
right-handed neutrino has non-vanishing complex Yukawa couplings when it 
unifies with the quarks at the GUT scale\footnote{For instance, we could 
think of a Georgi-Jarlskog vev distinguishing up-quark and neutrino Yukawas
and being zero for the neutrinos \cite{Georgi:1979df}.}. In this case the
neutrino decay through GUT Higgses could violate $CP$ and generate the 
observed baryon asymmetry. 

Finally we would like to point out another possibility where we introduce
more SM singlets mixed with the three ``standard'' right-handed neutrinos. An
example of this situation is provided by the so-called ``double seesaw''
\cite{Mohapatra:1986aw} 
mechanism. In this case, the right-handed Majorana masses are generated
through a second seesaw with these additional singlets. The singlets would 
decay only through its
mixings with the three right-handed neutrinos and, as we are introducing 
another free parameter, we can easily make any of these new singlets to 
dominate the energy density of the universe. Similarly, these singlets 
would easily generate the observed
baryon asymmetry through the usual neutrino Yukawa couplings. The problem
of how to generate a thermal abundance of these singlets could be solved again
if they are charged under a gauged B--L symmetry.

\section{Conclusions}
In this work, we have shown that an early epoch of matter 
domination by a long-lived massive particle can help to solve some of the 
problems of primordial inflation. We have seen that the large entropy 
production generated by the decays of such particle can dilute 
unwanted relics from higher temperatures, relaxing the  constraints 
on the inflationary reheating temperature. In supersymmetric theories this
mechanism can help to solve the gravitino problem. Moreover, a long period of 
matter domination reduces the number of $e$-foldings before the end of
inflation at which the observable cosmological perturbations were generated.
In the Standard Model a natural candidate for such heavy, long-lived particle 
is a heavy right-handed neutrino. For low enough mass of the lightest
left-handed neutrino and neutrino Yukawa mixings sufficiently close to the
PNMS mixing matrix, the right-handed neutrino dominates the energy density of
the universe for a long time and generates a large amount of entropy in its
decay.  In this case, we show that its decays can also generate the 
observed baryon asymmetry for right-handed neutrino masses well above the 
bound from gravitino overproduction.

\section*{Acknowledgments}
The authors are grateful to Scott Dodelson, Diego Saez and Lorenzo Sorbo
for useful discussions  and acknowledge support from the Spanish MEC and FEDER
under  Contract FPA2005-01678, European program MRTN-CT-2006-035482
``Flavianet''  and the Generalitat Valenciana under Contract GV05/267.


\begin{thebibliography}{99}
\bibitem{inflation}
A.~H.~Guth,
  Phys.\ Rev.\  D {\bf 23}, 347 (1981);\\
A.~D.~Linde,
  Phys.\ Lett.\  B {\bf 108}, 389 (1982); \\
For a recent review, see:  A.~D.~Linde,
  Phys.\ Rept.\  {\bf 333} (2000) 575.



\bibitem{Hinshaw:2008kr}
  G.~Hinshaw {\it et al.}  [WMAP Collaboration],
  arXiv:0803.0732 [astro-ph]; \\
  E.~Komatsu {\it et al.}  [WMAP Collaboration],
  arXiv:0803.0547 [astro-ph].


\bibitem{Alabidi:2005qi}
  L.~Alabidi and D.~H.~Lyth,
  JCAP {\bf 0605} (2006) 016
  [arXiv:astro-ph/0510441].


\bibitem{Nagano:1998aa}
  T.~Nagano and M.~Yamaguchi,
  Phys.\ Lett.\  B {\bf 438} (1998) 267
  [arXiv:hep-ph/9805204].

\bibitem{Kawasaki:1999na}
  M.~Kawasaki, K.~Kohri and N.~Sugiyama,
  Phys.\ Rev.\ Lett.\  {\bf 82} (1999) 4168
  [arXiv:astro-ph/9811437].


\bibitem{Kohri:2004qu}
  K.~Kohri, M.~Yamaguchi and J.~Yokoyama,
  Phys.\ Rev.\  D {\bf 70} (2004) 043522
  [arXiv:hep-ph/0403043].

\bibitem{Pradler:2006hh}
  J.~Pradler and F.~D.~Steffen,
  Phys.\ Lett.\  B {\bf 648} (2007) 224
  [arXiv:hep-ph/0612291].

\bibitem{Kitano:2008tk}
  R.~Kitano, H.~Murayama and M.~Ratz,
  Phys.\ Lett.\  B {\bf 669} (2008) 145
  [arXiv:0807.4313 [hep-ph]].


\bibitem{Lyth:1995hj}
  D.~H.~Lyth and E.~D.~Stewart,
  Phys.\ Rev.\ Lett.\  {\bf 75} (1995) 201
  [arXiv:hep-ph/9502417]; \\
 D.~H.~Lyth and E.~D.~Stewart,
  Phys.\ Rev.\  D {\bf 53} (1996) 1784
  [arXiv:hep-ph/9510204].

\bibitem{Kolb:1990vq}
  E.~W.~Kolb and M.~S.~Turner,
  ``The Early Universe,''
  Front.\ Phys.\  {\bf 69} (1990) 1.

\bibitem{Dodelson:2003ft}
  S.~Dodelson,
  ``Modern Cosmology,''
{\it  Amsterdam, Netherlands: Academic Pr. (2003) 440 p}

\bibitem{Turner:1984ff}
  M.~S.~Turner,
  Phys.\ Rev.\  D {\bf 31}, 1212 (1985).

\bibitem{Scherrer:1984fd}
  R.~J.~Scherrer and M.~S.~Turner,
  Phys.\ Rev.\  D {\bf 31}, 681 (1985).



\bibitem{Boubekeur}
L.~Boubekeur and P.~Creminelli,
  Phys.\ Rev.\  D {\bf 73}, 103516 (2006)
  [arXiv:hep-ph/0602052].


\bibitem{seesawrev}
  P.~Minkowski,
  Phys.\ Lett.\  B {\bf 67}, 421 (1977); \\
R.M.~Gell-Mann, P.~Ramond and R.~Slansky, in {\it Supergravity}, p.315,
ed. by F Nieuwenhuizen and D Friedman (North-Holland, Amsterdam, 1979);\\
T. Yanagida in Proc. {\it Workshop on the Unified Theory and 
the Baryon Number in the Universe}, ed. by O. Sawada and A. Sugamoto 
(KEK, Japan, 1979);\\
  R.~N.~Mohapatra and G.~Senjanovic,
  Phys.\ Rev.\  D {\bf 23}, 165 (1981).

\bibitem{neutrinorev}
  B.~Kayser,
  arXiv:0804.1497 [hep-ph].

\bibitem{Fischler:1990gn}
  W.~Fischler, G.~F.~Giudice, R.~G.~Leigh and S.~Paban,
  Phys.\ Lett.\  B {\bf 258}, 45 (1991).
\bibitem{Buchmuller:1992qc}
  W.~Buchmuller and T.~Yanagida,
  Phys.\ Lett.\  B {\bf 302}, 240 (1993).


\bibitem{Kawasaki:2004qu}
 M.~Kawasaki, K.~Kohri and T.~Moroi,
 Phys.\ Rev.\  D {\bf 71} (2005) 083502
 [arXiv:astro-ph/0408426].

\bibitem{Sato:1995du}
  J.~Sato,
  Phys.\ Rev.\  D {\bf 53} (1996) 3884
  [arXiv:hep-ph/9508269].

\bibitem{Aulakh:1999pz}
  C.~S.~Aulakh, B.~Bajc, A.~Melfo, A.~Rasin and G.~Senjanovic,
  Phys.\ Lett.\  B {\bf 460} (1999) 325
  [arXiv:hep-ph/9904352].

\bibitem{Aulakh:2000sn}
  C.~S.~Aulakh, B.~Bajc, A.~Melfo, A.~Rasin and G.~Senjanovic,
  Nucl.\ Phys.\  B {\bf 597} (2001) 89
  [arXiv:hep-ph/0004031].


\bibitem{baryo}
  A.~Riotto and M.~Trodden,
  Ann.\ Rev.\ Nucl.\ Part.\ Sci.\  {\bf 49} (1999) 35
  [arXiv:hep-ph/9901362]; \\
  W.~Buchmuller, R.~D.~Peccei and T.~Yanagida,
  Ann.\ Rev.\ Nucl.\ Part.\ Sci.\  {\bf 55} (2005) 311
  [arXiv:hep-ph/0502169]; \\
  S.~Davidson, E.~Nardi and Y.~Nir,
  arXiv:0802.2962 [hep-ph].


\bibitem{PDG}
  W.~M.~Yao {\it et al.}  [Particle Data Group],
  J.\ Phys.\ G {\bf 33} (2006) 1.

\bibitem{Masiero:2002jn}
  A.~Masiero, S.~K.~Vempati and O.~Vives,
  Nucl.\ Phys.\  B {\bf 649}, 189 (2003)
  [arXiv:hep-ph/0209303].


\bibitem{Georgi:1979df}
  H.~Georgi and C.~Jarlskog,
  Phys.\ Lett.\  B {\bf 86}, 297 (1979);\\
 G.~G.~Ross, L.~Velasco-Sevilla and O.~Vives,
  Nucl.\ Phys.\  B {\bf 692} (2004) 50
  [arXiv:hep-ph/0401064].


\bibitem{Mohapatra:1986aw}
  R.~N.~Mohapatra,
  Phys.\ Rev.\ Lett.\  {\bf 56}, 561 (1986);
\\
  R.~N.~Mohapatra and J.~W.~F.~Valle,
  Phys.\ Rev.\  D {\bf 34}, 1642 (1986).



\end{thebibliography}
\end{document}